\documentclass[11pt]{article}

\usepackage[margin=1in]{geometry}
\usepackage{graphicx}
\usepackage{booktabs}
\usepackage{amsmath}
\usepackage{amssymb}
\usepackage{siunitx}
\usepackage{url}
\usepackage[numbers,sort&compress]{natbib}
\usepackage{hyperref}
\hypersetup{
    colorlinks=true,
    linkcolor=black,
    citecolor=blue!60!black,
    urlcolor=blue!60!black,
}
\usepackage{caption}
\usepackage{microtype}
\usepackage{xcolor}

\usepackage{titlesec}
\titleformat*{\section}{\large\bfseries}
\titleformat*{\subsection}{\normalsize\bfseries}

\usepackage[htt]{hyphenat}

\title{%
    Why fragmented parliaments stop passing legislation: \\
    Opposition discipline and representation across four democratic institutions
}
\author{%
    Fuad Ali
}
\date{August 23, 2026}

\begin{document}
\maketitle

\begin{abstract}
\noindent
Parliamentary systems pass more bills than presidential systems at baseline,
but collapse to near-zero passage under party-system fragmentation. The
literature offers three competing micro-explanations: coalition-formation
failure, party discipline, and committee gatekeeping. These operate
simultaneously in any real legislature, so observational studies struggle to
separate their contributions. We present an agent-based model that compares
four democratic institutions: pure parliamentary, pure republican/presidential,
premier-presidential (France), and president-parliamentary (Russia). Across
four scenarios and $N=200$ seeds per cell we report bootstrap confidence
intervals, Morris screening, Sobol variance decomposition, mechanism
ablations, and a hung-parliament variant comparison. Three findings emerge.
First, government formation failure alone does not halt legislation: when a
fragmented parliament reverts to personal voting, parliamentary passage
($46.4\%$) is statistically indistinguishable from the presidential benchmark
($44.8\%$); collapse requires cohesive opposition obstruction, which drives
passage to $0.05\%$. Second, disabling discipline restores fragmented passage to
$46.7\%$, and the rescue magnitude is monotone across the four institutions
in a pattern that survives varying the common discipline level. Third, the
passage--representation tradeoff is a single spectrum: parliamentary
maximises throughput at the cost of representational fidelity; republican
maximises fidelity via the presidential veto; semi-presidential variants
split the difference.

\medskip
\noindent\textbf{Keywords:} legislative studies, institutional design,
party discipline, opposition cohesion, minority government,
semi-presidentialism, agent-based modelling, Morris screening, Sobol
sensitivity.
\end{abstract}

\begin{center}
    \refstepcounter{figure}
    \label{fig:forest}
    \includegraphics[width=0.88\linewidth]{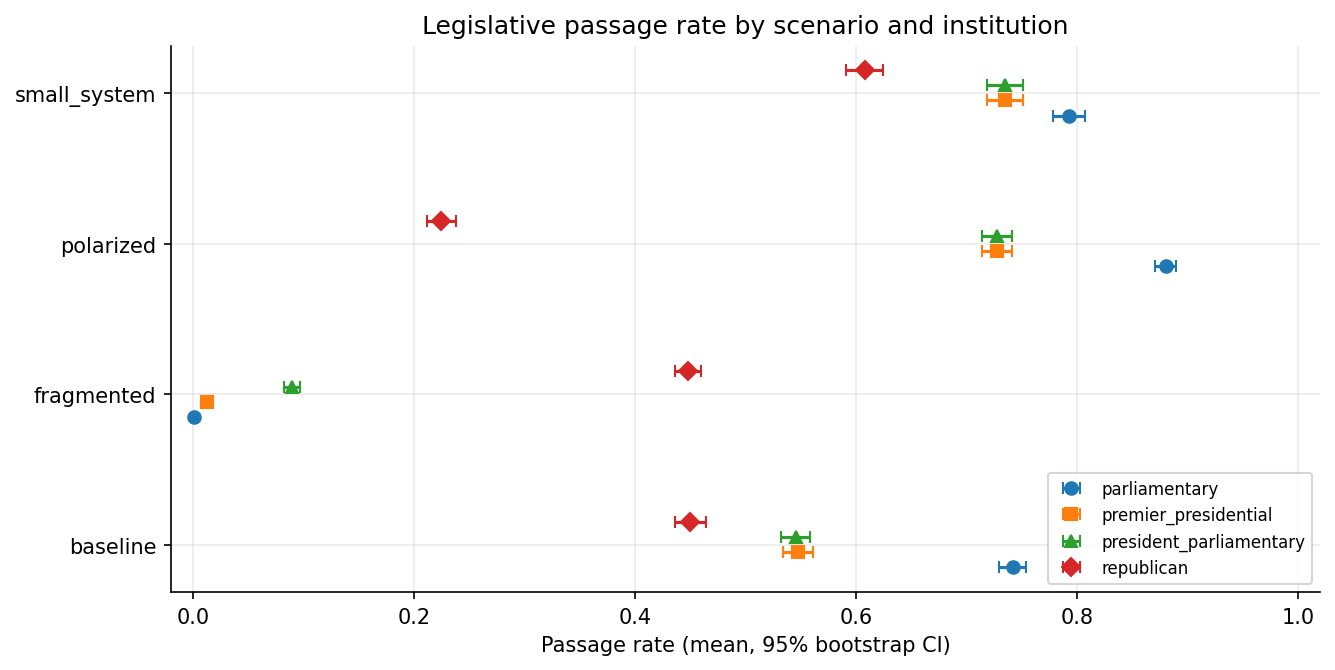}\\[4pt]
    \begin{minipage}{0.88\linewidth}
    \small
    \textbf{Figure~\thefigure:}~Mean passage rate with $95\%$ bootstrap
    confidence intervals across four institutions and four scenarios
    ($N=200$ seeds per cell). The baseline parliamentary advantage reverses
    under fragmentation: pure parliamentary and premier-presidential
    collapse to near-zero passage, while president-parliamentary partially
    rescues and republican is unaffected. Section~\ref{sec:results:passage}
    reports the full passage-rate table;
    Section~\ref{sec:results:collapse} decomposes the collapse into
    government-formation failure versus opposition-side cohesion and shows
    only the latter is causal;
    Section~\ref{sec:results:ablation} locates the mechanism in party
    discipline.
    \end{minipage}
\end{center}

\section{Introduction}
\label{sec:intro}

Democracies differ dramatically in how much legislation they produce. In the
United Kingdom, government bills almost never fail on the floor of the House
of Commons; the last government bill defeated at second reading was the
Shops Bill of 1986 \citep{shopsbill1986}. Germany passes roughly $90\%$ of
government-initiated bills \citep{dalton1993}. In contrast, only $3$--$7\%$
of bills introduced to the U.S. Congress are enacted into law, though most of
that figure reflects filtering at committee stage rather than genuine floor
rejection \citep{govtrack2023}. France, a premier-presidential system,
adopted roughly three-fifths of its laws in the most recent legislature as
government-sponsored \emph{projets de loi} \citep{anstats2024}. Russia's
State Duma, a president-parliamentary system, passes presidential bills at
near-universal rates (a perfect success record realised in the $2007$--$11$
convocation), while backbench initiatives rarely become law at all
\citep{noble2018}.

These differences are hard to attribute to demand: mass policy preferences
in established democracies, while not identical, overlap heavily and cluster
near the centre of the left--right scale \citep{powell2000, caughey2019},
yet the legislatures elected by these publics differ by orders of magnitude in
passage rate. The natural explanation is the rules. But
the rules interact in ways that are difficult to disentangle. Three
mechanistic accounts compete to explain why some legislatures produce laws
efficiently and others grind to a halt. The first emphasises
\textbf{coalition formation}: where a single party cannot hold a working
majority, bargaining over portfolios and policy may fail to produce a
viable government, blocking any legislative action before it starts
\citep{strom1990, laver1990}. The second emphasises \textbf{party
discipline}: strong whipping aligns MPs with their leadership rather than
their personal or constituent preferences, enabling the government to pass
bills that a free vote would reject \citep{bowler1999, cox1993}. The third
emphasises \textbf{committee gatekeeping}: structural rules that allow
small numbers of legislators to kill bills before floor consideration
\citep{shepsle1978, tsebelis2002}. These accounts are not mutually
exclusive, but their relative contributions to any observed passage pattern
are difficult to estimate because all three operate simultaneously in
every real legislature.

This paper uses an agent-based model (ABM) to attribute passage outcomes to
specific mechanisms. An ABM can implement coalition formation, party
discipline, and committee gatekeeping in a single deterministic simulation
and disable them one at a time while holding everything else fixed. The
mechanism-level counterfactuals required for causal attribution are
straightforward in a simulation and impossible in observational data.
Prior ABM work in legislative studies has typically modelled a single
mechanism at a time: coalition formation \citep{laver1990},
electoral mandate structure \citep{fowler2007}, and spatial voting in
the \citet{downs1957} tradition. Our contribution is to integrate the
three competing mechanisms in one framework and compare their
contributions across institutions and scenarios, building on the stylised
pattern that presidential systems achieve lower legislative success rates
than prime-ministerial systems while remaining robust to coalition failure
\citep{shugart2008, saiegh2014}.

We compare four institutional types. Pure \textbf{parliamentary} systems
follow the Westminster-Continental European template: the cabinet is drawn
from the legislature and falls on failed confidence votes. Pure
\textbf{republican} (presidential) systems follow the U.S. template: the
executive is elected separately, has formal veto power, and is not
responsible to the legislature. Between these two sit the
\textbf{semi-presidential} variants classified by \citet{shugart1992}.
\textbf{Premier-presidential} systems (France, Portugal, Ireland) have a
directly elected president alongside a cabinet responsible to parliament,
but the cabinet is formed by parliamentary majority and the president cannot
unilaterally dismiss the prime minister. \textbf{President-parliamentary}
systems (Russia, Weimar Germany) have the same pair of offices but the
president forms the government and can dismiss the PM without a parliamentary
vote. In our model both variants are reachable from a single class through
two independent configuration toggles.

Across these four institutions we run $N=200$ random seeds per scenario,
report bootstrap $95\%$ confidence intervals, use Morris elementary-effects
screening and Sobol variance decomposition to identify the dominant
parameters, and run mechanism ablations (disabling committees, discipline,
or veto one at a time) to attribute passage contributions. We also check
the robustness of the main ordering by varying the common discipline level
applied to all institutions.

Three findings structure our contribution. First (\S\ref{sec:results:collapse}),
government formation failure alone does not halt legislation. We compare two
defensible readings of a hung parliament: one in which cohesive opposition
blocks every bill, and a personal-vote null in which MPs revert to their own
preferences when no whip is in play. Under the null, fragmented-parliamentary
passage ($46.4\%$) matches the presidential benchmark ($44.8\%$); under
obstruction it collapses to $0.05\%$. Cohesion on the opposition side is
what converts formation failure into collapse. Second
(\S\ref{sec:results:ablation}), the no-discipline ablation restores parliamentary
passage from $0.05\%$ to $46.7\%$, and the rescue magnitude is monotone
across the four institutions in the order parliamentary $>$
premier-presidential $>$ president-parliamentary $>$ republican, an ordering
that holds at all tested common discipline levels and reverses sign under
polarisation: disabling discipline there costs parliamentary $66$ percentage
points. Discipline aggregates votes wherever a coalition exists; its absence
or presence on the \emph{opposition} side decides whether a hung parliament
legislates at all. Third (\S\ref{sec:results:tradeoff}), the passage--representation
tradeoff is a single spectrum. Parliamentary passes the most bills but its
passed-bill ideology matches no particular filter (policy representation gap
$\approx +0.04$ at polarised). Republican filters aggressively through the
presidential veto (gap $\approx -0.53$ at polarised) at the cost of high
non-passage rates. Semi-presidential variants sit between.

The remainder of the paper proceeds as follows.
Section~\ref{sec:model} describes the model;
Section~\ref{sec:design} describes the experimental design;
Section~\ref{sec:results} presents the results;
Section~\ref{sec:discussion} discusses implications and relates the
findings to prior work; Section~\ref{sec:limitations} states limitations;
Section~\ref{sec:reproducibility} documents code, data, and the
interactive companion app.

\section{Model description}
\label{sec:model}

\subsection{Agents and state}
The model contains four agent types:
\textbf{constituencies} with a fixed two-dimensional ideology (economic
$\times$ social) and a population;
\textbf{parties} with an ideology and a name;
\textbf{legislators} with an ideology, an assigned constituency, and an
assigned party; and
\textbf{committees} with a policy-area jurisdiction (an ideology centre
and radius) and party-proportional membership.
Bills are ephemeral objects with an ideology and a salience, instantiated
per legislative step.

All four institutions share the same agent types and initialisation
procedure. They differ only in the bill-processing pipeline invoked by
\texttt{pass\_legislation(bill)} and in the institution-specific state that
pipeline reads and writes (government, cabinet, president, confidence votes,
vetoes, dismissals).

Initialisation is deterministic. Constituencies, parties, and legislators
each receive evenly spaced ideologies along the main diagonal of the space,
from $(-1,-1)$ to $(1,1)$; legislator $i$ is assigned to party $i \bmod
n_{\text{parties}}$ and constituency $i \bmod n_{\text{constituencies}}$.
The round-robin party assignment means every party's members span the full
ideological range: parties are maximally heterogeneous internally, so the
whip has real aggregation work to do. There are no stochastic initial
conditions; the seed drives only process randomness (bill draws, vote
rolls, committee sampling). Section~\ref{sec:results:clustered} reports a
robustness check that replaces this spread initialisation with
party-clustered legislator ideologies.

\subsection{Passage pipelines}
Parliamentary: bills route to a committee with matching jurisdiction
(committees can approve, amend, or kill); with probability
$\texttt{confidence\_matter\_rate}$ a bill is treated as a confidence vote
and whipped harder; a floor vote with party discipline decides passage;
failed confidence votes trigger government collapse and reformation.

When no governing coalition can be formed at all (the fragmented scenario
makes this certain for majority-driven institutions), the whip structure is
undefined, and two readings of the situation are possible:
\begin{itemize}
    \item \textbf{cohesive obstruction}: every MP counts as ``opposition''
        and votes against bills with probability
        $\texttt{discipline} \times \texttt{multiplier}$, an anti-system
        reading in which polarised blocs block all legislation
        \citep{sartori1976, capoccia2005};
    \item \textbf{personal vote}: no whip is in play, so every MP reverts to
        their own preference, the issue-by-issue-majority world of
        minority-government governance \citep{strom1984, strom1990,
        laver1990}.
\end{itemize}
A configuration flag, $\texttt{hung\_parliament\_behavior} \in
\{\texttt{cohesive\_obstruction}, \texttt{personal\_vote}\}$, selects between
them; the shipped default is cohesive obstruction. Section~\ref{sec:results:collapse}
shows that this single choice determines whether fragmentation produces a
collapse at all, which is why we treat it as an experimental factor rather
than a silent implementation detail.

Republican: bills route to a committee (with stronger gatekeeping power);
a floor vote with weak discipline decides legislative passage; the
executive vetoes probabilistically by ideological distance from the bill;
a two-thirds override attempt follows vetoes.

Semi-presidential: hybridises both pipelines. Committees route bills;
bills may be treated as confidence matters; a floor vote with moderate
discipline decides legislative passage; the directly-elected president
then checks a veto; if the president can dismiss the PM
(president-parliamentary variant) the dismissal check fires on every bill
and may collapse the government independently of confidence votes.

\subsection{Configuration toggles for semi-presidential variants}
Two \texttt{SemiPresidentialConfig} toggles distinguish the variants:

\begin{itemize}
    \item \texttt{government\_formation} $\in$ \{\texttt{majority\_driven},
        \texttt{president\_driven}\}. Majority-driven formation follows the
        parliamentary rule (largest party alone if it holds a majority,
        else two-party coalition). President-driven formation appoints the
        president's party as the governing party, adding a second party for
        majority only if available; otherwise a minority cabinet.
    \item \texttt{president\_can\_dismiss\_pm} $\in$ \{\texttt{False},
        \texttt{True}\}. When \texttt{True}, the president dismisses the PM
        with probability \texttt{presidential\_dismissal\_rate} whenever
        their ideological distance exceeds $0.5$.
\end{itemize}

Premier-presidential corresponds to
(\texttt{majority\_driven}, \texttt{False}); president-parliamentary to
(\texttt{president\_driven}, \texttt{True}). Non-canonical combinations are
reachable but not the focus here.

\subsection{ODD protocol}
A full ODD description \citep{grimm2010, grimm2020} appears in
Appendix~\ref{app:odd}. All agent states, scheduling, design concepts,
initialisation, and submodel equations are documented there.

\section{Experimental design}
\label{sec:design}

\subsection{Scenarios}
Four scenarios vary the political environment while holding institutional
rules fixed:
\textbf{baseline} ($20$ legislators, $3$ parties, $6$ constituencies, $25$
bills, bill ideology range $[-1, 1]$);
\textbf{fragmented} ($24$ legislators, $5$ parties, $8$ constituencies, $30$
bills, range $[-1, 1]$);
\textbf{polarised} ($16$ legislators, $2$ parties, $4$ constituencies, $20$
bills, range $[-1.5, 1.5]$); and
\textbf{small-system} ($10$ legislators, $2$ parties, $3$ constituencies,
$15$ bills, range $[-0.8, 0.8]$).

\subsection{Statistical harness}
Each (scenario, institution) cell is replicated over $N=200$ seeds. We
report means with percentile bootstrap $95\%$ confidence intervals
(\texttt{scipy.stats.bootstrap}, $9\,999$ resamples). Pairwise institution
comparisons use Welch's $t$-test and Mann-Whitney $U$; effect sizes are
reported as Cohen's $d$. When zero-variance cells produce undefined
$p$-values, those are recorded as \texttt{NaN} rather than imputed.

\subsection{Sensitivity analysis}
For each institution we define a six-parameter problem and run Morris
elementary-effects screening ($r=20$ trajectories, $4$ levels) and Sobol
variance decomposition ($N=256$ base samples, first-order and total-order
indices). Three seeds per design point reduce stochastic variance without
inflating cost. The six parameters per institution always include
\texttt{discipline\_strength}, \texttt{committee\_gatekeeping\_power},
\texttt{num\_parties}, and \texttt{num\_constituencies}; the remaining two
are institution-specific.

\subsection{Mechanism ablations}
Three ablations modify the running model without touching the production
code. \texttt{no\_committees} replaces the committee-routing function with
a pass-through; \texttt{no\_discipline} sets
\texttt{discipline\_strength = 0}; \texttt{no\_veto} (where applicable)
replaces the executive-veto check with \texttt{False}. Each ablation is
monkey-patched on the live model instance; no class-level mutation occurs.

\subsection{Discipline-default robustness check}
To test whether the observed ordering depends on our default
discipline values, we sweep a common discipline level
$D \in \{0.0, 0.1, \ldots, 0.9\}$ applied simultaneously to all four
institutions, and compute
$\mathrm{rescue}(D) = \mathrm{passage}(D=0) - \mathrm{passage}(D)$ per
institution at $N=100$ seeds per cell.

\subsection{Hung-parliament variant comparison}
\label{sec:design:hung}
The fragmentation scenario makes coalition formation impossible for the
majority-driven institutions (largest party $\approx 21\%$ of seats; two-party
combinations also fall short), so the whip structure defined in
\S\ref{sec:model} never engages. We run the full $4 \times 4$ grid at
$N=200$ seeds under both values of
$\texttt{hung\_parliament\_behavior}$. Because the flag can only bind when
the coalition list is empty, the two variants are bit-identical in the
baseline, polarised, and small-system scenarios (where governments form) and
for republican (no formation gate) and president-parliamentary (the president
always seats at least a minority cabinet); all differences are therefore
concentrated in the fragmented cells for parliamentary and
premier-presidential. The runner uses the same bill-generation order as the
main harness, so default-variant rows reproduce the main tables exactly.

\section{Results}
\label{sec:results}

\subsection{Passage rates across scenarios}
\label{sec:results:passage}

Figure~\ref{fig:forest} (placed on the title page) summarises per-scenario
mean passage rates with $95\%$ bootstrap CIs for all four institutions. At
baseline the expected ordering holds: parliamentary $0.742$
[$0.729$, $0.754$], premier-presidential $0.547$, president-parliamentary
$0.546$, republican $0.450$ [$0.436$, $0.464$]. Parliamentary's advantage
$(+29\text{pp})$ over republican is highly significant (Welch's $t = 31.1$,
$p < 10^{-100}$; Cohen's $d = 3.11$).

\subsection{The fragmentation collapse decomposed}
\label{sec:results:collapse}

\begin{figure}[ht]
    \centering
    \includegraphics[width=0.95\linewidth]{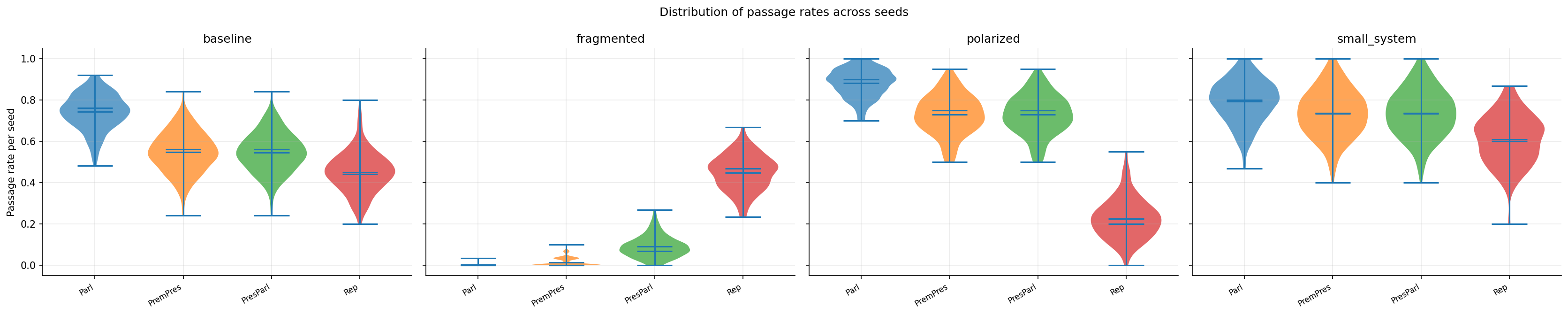}
    \caption{Per-seed passage-rate distributions, faceted by scenario. The
    fragmented cell shows near-zero passage for both majority-driven
    institutions (parliamentary, premier-presidential) and a partial rescue
    under president-parliamentary.}
    \label{fig:violin}
\end{figure}

Under the shipped default (cohesive obstruction), fragmentation collapses
majority-driven institutions: parliamentary passage falls to $0.0005$
($95\%$ CI $[0, 0.0012]$), premier-presidential to $0.013$, while
president-parliamentary partially rescues passage to $0.086$ through its
president-driven minority-cabinet rule and republican holds at $0.448$
(Figure~\ref{fig:violin}; the collapse is a regime property, not a tail
event).

\begin{figure}[ht]
    \centering
    \includegraphics[width=0.98\linewidth]{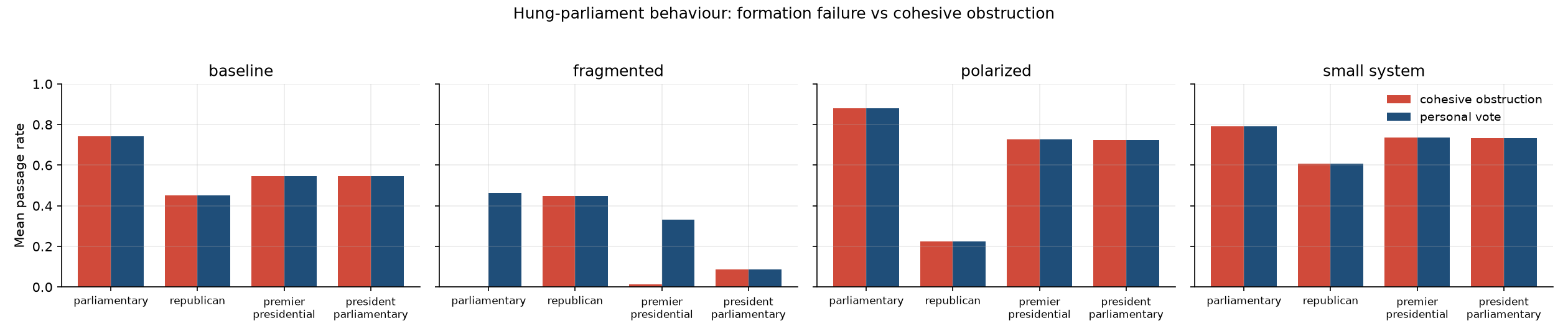}
    \caption{Mean passage rate by institution and hung-parliament behaviour,
    per scenario ($N=200$ seeds). Variants differ only in the fragmented
    cells of the majority-driven institutions: formation failure alone leaves
    parliamentary passage at presidential levels ($46.4\%$ vs.\ republican's
    $44.8\%$); cohesive opposition obstruction produces the collapse.}
    \label{fig:hung}
\end{figure}

But which ingredient produces the collapse? Formation failure
and whip behaviour are confounded in the default runs, because both are
absent simultaneously. Varying
$\texttt{hung\_parliament\_behavior}$ unconfounds them
(\S\ref{sec:design:hung}). Table~\ref{tab:hung} and
Figure~\ref{fig:hung} show the result: under the personal-vote null,
fragmented parliamentary passage \emph{recovers} to $46.4\%$
$[45.3, 47.6]$, statistically indistinguishable from the presidential
benchmark of $44.8\%$ $[43.6, 46.0]$, while under cohesive obstruction it
stays collapsed at $0.05\%$. Premier-presidential moves from $1.3\%$ to
$33.3\%$ under the null, below parliamentary because the presidential veto
filters bills even without obstruction. Republican and
president-parliamentary are identical across variants, as required: neither
can enter a hung state.

\begin{table}[ht]
    \centering
    \caption{Passage rate under fragmentation by hung-parliament behaviour
    ($N=200$ seeds per cell). The flag binds only where the coalition list
    can be empty.}
    \begin{tabular}{lccc}
        \toprule
        Institution & Cohesive obstruction & Personal vote & $\Delta$ \\
        \midrule
        parliamentary & $0.0005$ & $0.464$ & $+0.464$ \\
        premier\_presidential & $0.013$ & $0.333$ & $+0.320$ \\
        president\_parliamentary & $0.086$ & $0.086$ & $0$ \\
        republican & $0.448$ & $0.448$ & $0$ \\
        \bottomrule
    \end{tabular}
    \label{tab:hung}
\end{table}

The decomposition reverses the causal story told by observational
comparisons. Coalition-formation failure, emphasised in the government-formation
literature \citep{strom1990, laver1990}, is \emph{not sufficient}: a
fragmented chamber whose members vote their preferences legislates at full
presidential-system rates. What generates the near-zero regime is cohesive
obstruction on the opposition side: the anti-system pattern in which blocs
whip against all business regardless of content \citep{sartori1976,
capoccia2005}. This matches the empirical record: fragmented parliaments with
weakly disciplined or office-seeking oppositions keep passing legislation
through minority and caretaker cabinets \citep{strom1984, laver1990}, while
polarised anti-system blocs produce legislative deadlock \citep{capoccia2005}.

\subsection{Sensitivity: different institutions, different bottlenecks}
\label{sec:results:sensitivity}

\begin{figure}[ht]
    \centering
    \includegraphics[width=0.95\linewidth]{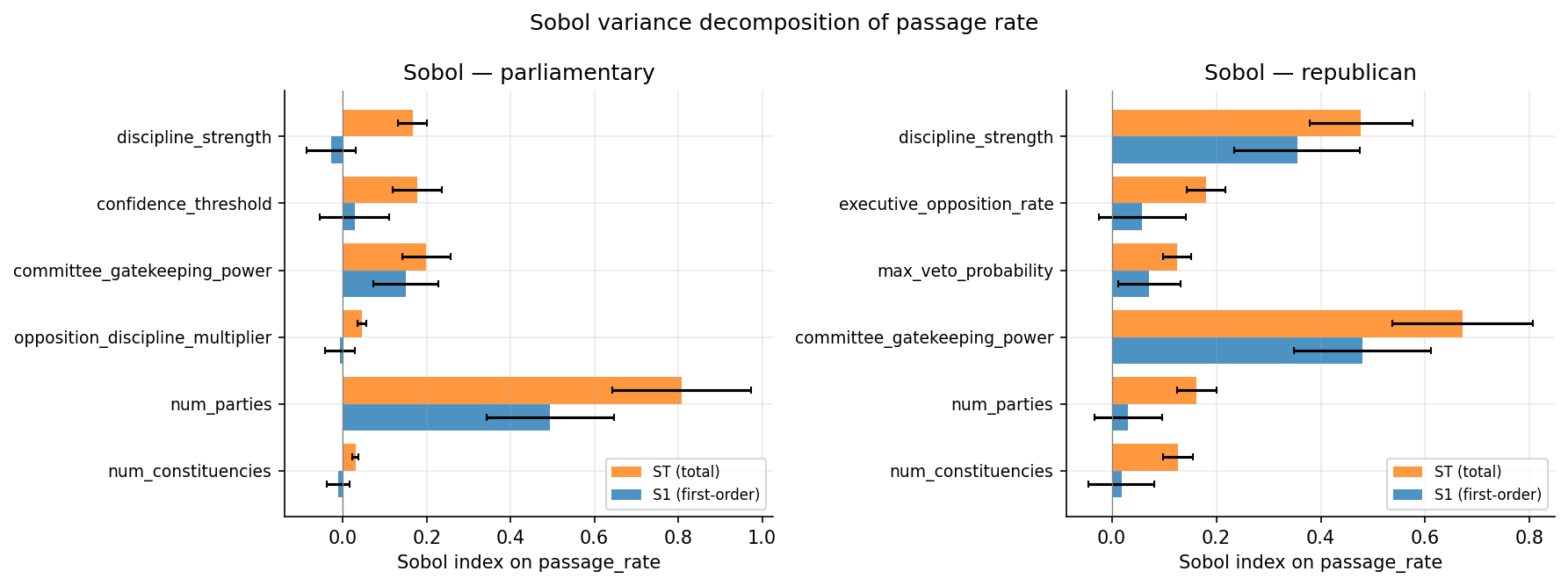}
    \caption{Sobol first-order ($S_1$) and total-order ($S_T$) indices for
    passage rate at baseline.}
    \label{fig:sobol}
\end{figure}

Sobol total-order indices at baseline (Fig.~\ref{fig:sobol}) show that
parliamentary passage is dominated by \texttt{num\_parties}
($S_T = 0.808$), consistent with the fragmentation-collapse finding.
Republican passage is dominated by
\texttt{committee\_gatekeeping\_power} ($S_T = 0.671$) with
\texttt{discipline\_strength} second ($S_T = 0.477$). The two institutions
have structurally different bottlenecks. Semi-presidential variants track
parliamentary in their dependence on \texttt{num\_parties}.

The same analysis on the fragmented scenario confirms the 	exttt{num\_parties}
dominance across all three majority-dependent institutions
($S_T \approx 0.76$--$0.79$). \texttt{confidence\_threshold} gains
importance under fragmentation ($+0.12$ in $S_T$ for parliamentary),
which is consistent with a binding majority-threshold interpretation.
Republican remains structurally insensitive to \texttt{num\_parties}
($S_T = 0.14$).

\subsection{Mechanism attribution via ablation}
\label{sec:results:ablation}

\begin{figure}[ht]
    \centering
    \includegraphics[width=0.85\linewidth]{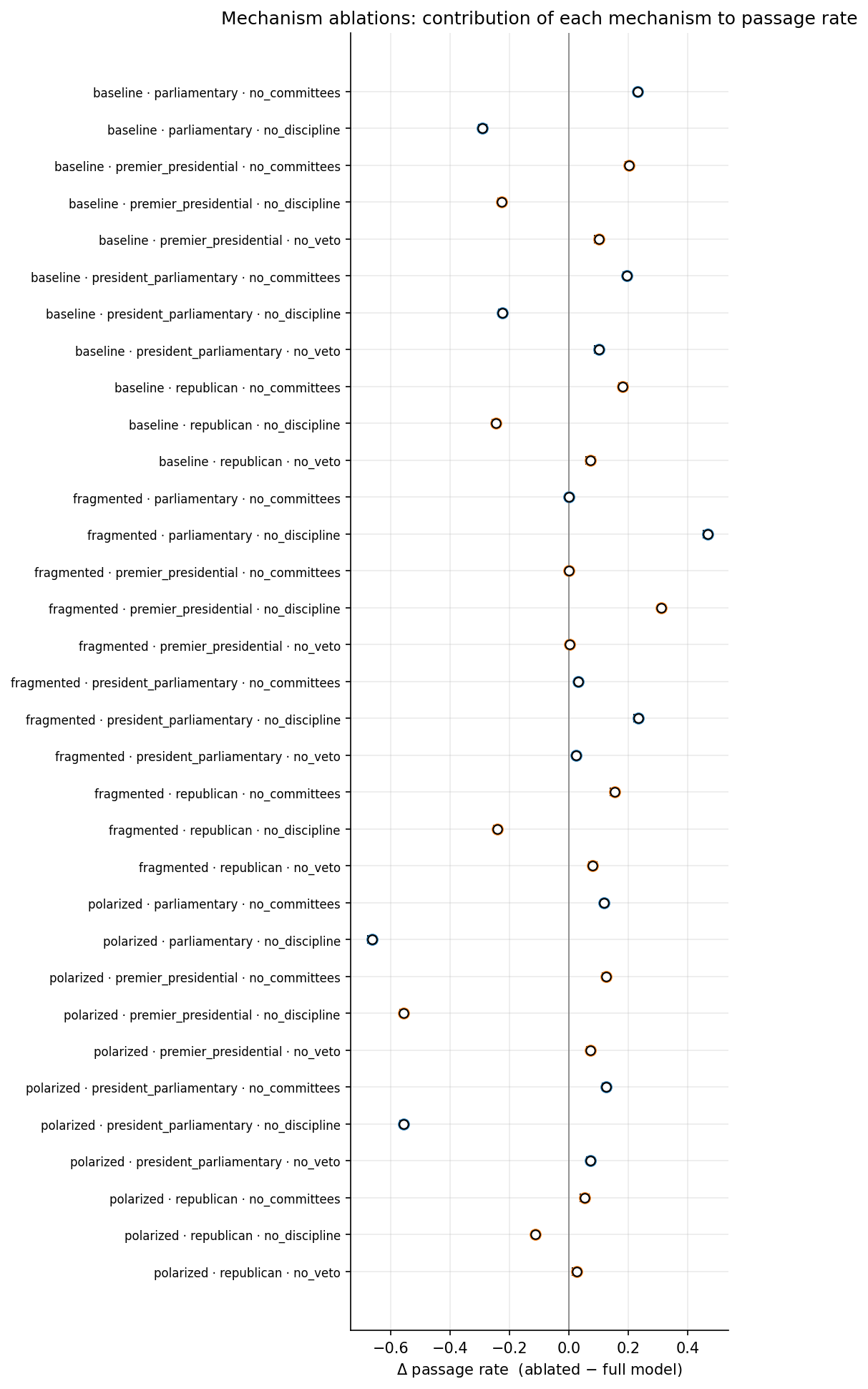}
    \caption{Ablation deltas ($\Delta$ passage rate relative to full model)
    for each (scenario, institution, ablation) cell at $N=200$ seeds.
    Confidence bars are $\pm 1.96\,\mathrm{SE}$.}
    \label{fig:ablation}
\end{figure}

Figure~\ref{fig:ablation} shows $\Delta\text{passage\_rate}$ relative to
the full model for each (scenario, institution, ablation) cell at $N=200$.
The headline result appears at fragmented:

\begin{table}[ht]
    \centering
    \caption{\texttt{no\_discipline} ablation $\Delta$ under fragmentation.}
    \begin{tabular}{lcc}
        \toprule
        Institution & Full-model passage & $\Delta$ (no\_discipline $-$ full) \\
        \midrule
        parliamentary & $0.0005$ & $+0.466$ \\
        premier\_presidential & $0.013$ & $+0.312$ \\
        president\_parliamentary & $0.086$ & $+0.228$ \\
        republican & $0.448$ & $-0.242$ \\
        \bottomrule
    \end{tabular}
    \label{tab:ablation-fragmented}
\end{table}

Disabling party discipline restores parliamentary passage from
near-zero to nearly $47\%$, closely matching the personal-vote hung
parliament of \S\ref{sec:results:collapse} ($46.4\%$) via two independent
manipulations (zeroing the whip everywhere vs.\ removing it only where no
government exists), which together cross-validate the attribution. The rescue
magnitude decreases monotonically across the four institutions in the order
parliamentary $>$ premier-presidential $>$ president-parliamentary $>$
republican. Republican is the only institution where zeroing discipline
\emph{hurts} rather than helps: without a majority-formation gate, discipline plays a different
role (crude vote aggregation on the floor) and removing it simply damages
the signal.

Under polarised scenarios the same ordering holds in magnitude but with
opposite sign (Table~\ref{tab:ablation-polarized}). Discipline \emph{enables}
passage when the coalition wants extreme bills to pass; disabling
discipline costs parliamentary $66$ percentage points. Republican, with
no coalition to whip, loses only $11$ percentage points. Across scenarios,
discipline is the mechanism majority-dependent institutions use to
\emph{aggregate votes}, rather than a mechanism that specifically causes
fragmentation pathologies.

\begin{table}[ht]
    \centering
    \caption{\texttt{no\_discipline} ablation $\Delta$ under polarisation.}
    \begin{tabular}{lcc}
        \toprule
        Institution & Full-model passage & $\Delta$ (no\_discipline $-$ full) \\
        \midrule
        parliamentary & $0.880$ & $-0.664$ \\
        premier\_presidential & $0.728$ & $-0.557$ \\
        president\_parliamentary & $0.723$ & $-0.554$ \\
        republican & $0.225$ & $-0.113$ \\
        \bottomrule
    \end{tabular}
    \label{tab:ablation-polarized}
\end{table}

\subsection{Passage--representation tradeoff}
\label{sec:results:tradeoff}

\begin{figure}[ht]
    \centering
    \includegraphics[width=0.98\linewidth]{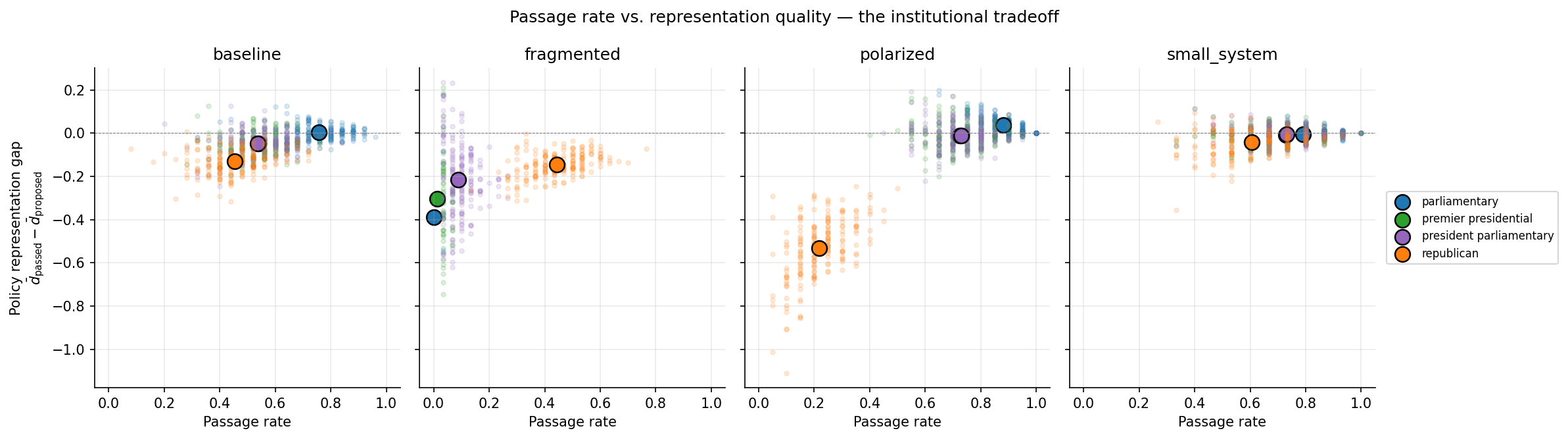}
    \caption{Passage rate versus policy representation gap, per scenario.
    Each large marker is the $N=200$ mean; the cloud is per-seed. Under
    polarisation the tradeoff is most visible: parliamentary at
    $(0.88, +0.04)$ vs.\ republican at $(0.22, -0.53)$.}
    \label{fig:tradeoff}
\end{figure}

Figure~\ref{fig:tradeoff} plots passage rate against a new metric, the
policy representation gap:
\begin{equation}
    \mathrm{gap} \;=\; \overline{d_{\text{passed}}}
                     - \overline{d_{\text{proposed}}}
\end{equation}
where $d$ is the L2 distance between a bill's ideology and the
constituency-median ideology. Negative gaps indicate the institution
filters passed bills toward the median; positive gaps indicate the
institution spreads passed bills relative to random proposal draws.

The tradeoff is sharpest under polarised scenarios:

\begin{table}[ht]
    \centering
    \caption{Passage--representation tradeoff at polarised ($N=200$).}
    \begin{tabular}{lcc}
        \toprule
        Institution & Passage rate & Representation gap \\
        \midrule
        parliamentary & $0.882$ & $+0.037$ \\
        premier\_presidential & $0.725$ & $-0.012$ \\
        president\_parliamentary & $0.730$ & $-0.010$ \\
        republican & $0.218$ & $-0.530$ \\
        \bottomrule
    \end{tabular}
    \label{tab:tradeoff-polarized}
\end{table}

Parliamentary passes $88\%$ of bills with essentially no ideological filter;
republican passes only $22\%$ but the bills that do pass are strongly
biased toward the constituent median, reflecting the presidential veto's
role as an ideological screen. Semi-presidential variants sit on the same
spectrum between them. Filtering and throughput are mechanistically
linked: any mechanism that increases one reduces the other.

\subsection{Robustness of the monotone ordering}

\begin{figure}[ht]
    \centering
    \includegraphics[width=0.98\linewidth]{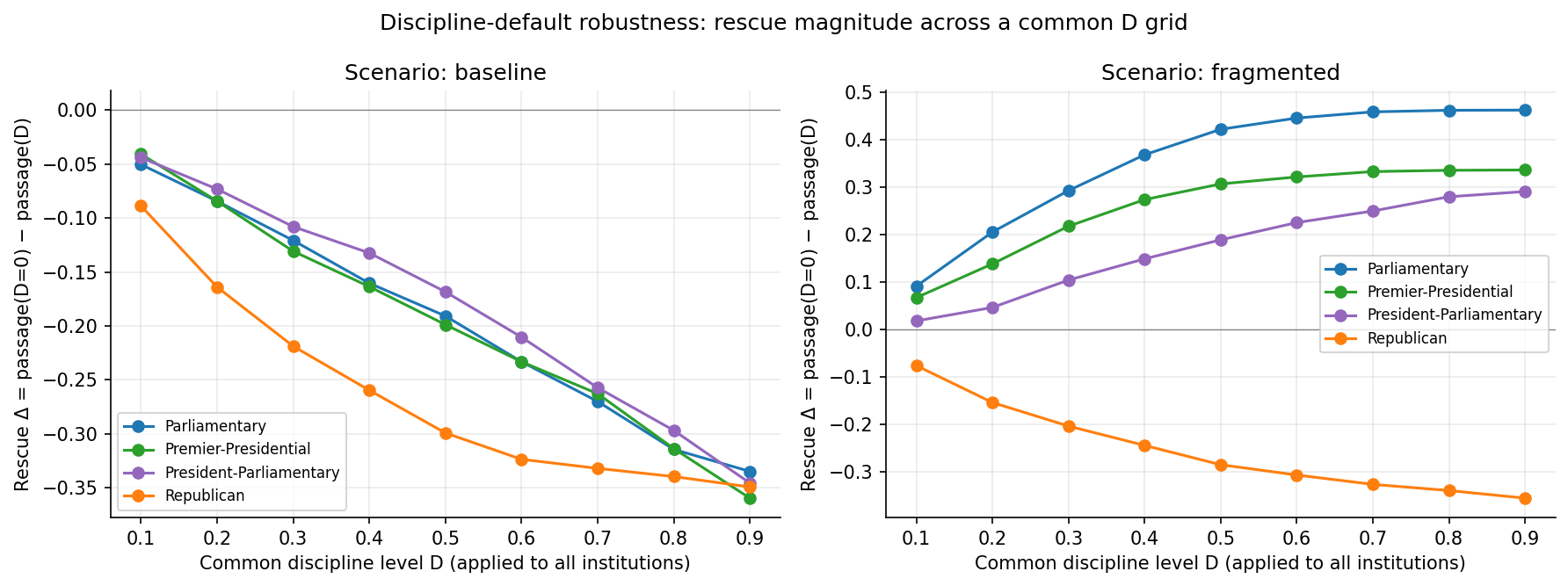}
    \caption{Rescue magnitude versus common discipline level $D$, per
    institution, at baseline and fragmented scenarios. Under fragmented
    (right panel), the monotone ordering parliamentary $>$
    premier-presidential $>$ president-parliamentary $>$ republican holds
    at all nine non-zero levels ($D=0$ is trivial by definition).}
    \label{fig:robustness}
\end{figure}

The ablation ordering in Table~\ref{tab:ablation-fragmented} could in
principle be an artefact of our default \texttt{discipline\_strength}
values, which happen to be ordered the same way across institutions
($0.8$, $0.6$, $0.6$, $0.4$). Figure~\ref{fig:robustness} tests this by
setting all four institutions' \texttt{discipline\_strength} to the same
value at each of the sweep's nine non-zero levels
($D \in \{0.1, \ldots, 0.9\}$) and measuring the rescue magnitude.

The monotone ordering parliamentary $>$ premier-presidential $>$
president-parliamentary $>$ republican holds at $9/9$ non-zero levels
under fragmentation. The structural claim survives: the ordering is not a
function of our chosen defaults. A regression test in the repository
enforces this at $D \in \{0.0, 0.2, 0.5, 0.8\}$ so that future mechanism
changes that break the ordering fail CI.

\label{sec:results:clustered}
A second initialisation concern is the spread assignment itself
(\S\ref{sec:model}): evenly spaced legislators with round-robin party
membership make parties maximally heterogeneous internally, which plausibly
inflates the \emph{magnitude} of the discipline effect: the whip has more
work to do when members span the whole spectrum. We therefore re-initialise
legislator ideologies as their own party's position plus Gaussian noise
($\sigma = 0.15$) and re-run the discipline ablation ($N=100$ seeds per
cell). Both findings keep their structure. Under fragmentation the rescue
ordering survives: parliamentary $+0.35$, premier-presidential $+0.23$,
president-parliamentary $+0.16$, republican $\approx 0$. Under polarisation
the sign flip survives: parliamentary $-0.74$, semi-presidential variants
near $-0.57$, republican flat. Magnitudes attenuate relative to the spread
initialisation (parliamentary's fragmented rescue falls from $+0.47$ to
$+0.35$), exactly as the heterogeneity account predicts. Within-party
heterogeneity scales the size of the discipline effect but not its ordering
or its sign structure.

\section{Discussion}
\label{sec:discussion}

\subsection{Discipline as a universal aggregation mechanism}

The hung-parliament decomposition (\S\ref{sec:results:collapse}) and the
ablation results across fragmented and polarised scenarios
(Tables~\ref{tab:ablation-fragmented} and \ref{tab:ablation-polarized})
reveal that party discipline is not simply a ``collapse trigger'' under
fragmentation or a ``passage enabler'' under polarisation. It is the same
mechanism in both cases: discipline aggregates legislators toward a bloc
position rather than the free-vote median. When that bloc is a governing
coalition near the floor median, discipline enforces passage. When it is an
opposition bloc facing a chamber with no government, the same aggregation
produces uniform obstruction. And when no bloc exists at all (the
personal-vote null), a fragmented chamber legislates at full presidential
rates, so formation failure alone blocks nothing.

The monotone ablation ordering across the four institutions reflects how
much each relies on a formation-gated coalition to act. Parliamentary requires a
majority coalition before any legislation is possible; premier-presidential
requires the same, but with a presidential veto adding a second gate;
president-parliamentary also nominally requires a majority, but the
president-driven formation rule lets the government function as a minority
cabinet when no majority forms; republican has no formation gate at all.
The rescue magnitude under \texttt{no\_discipline} scales with this
dependence. This reading is consistent with the demonstration by
\citet{strom1984, strom1990} that minority governments are rational, functioning cabinet
solutions rather than instability artefacts (our personal-vote variant is a
stylised version of their issue-by-issue-majority world) and with
the \citet{shugart1992} typology of semi-presidential variants.

The obstruction side of the decomposition has older roots.
The anti-system parties of \citet{sartori1976} (blocs defined by opposition to
the democratic rules of the game rather than by policy competition)
generate exactly the uniform against-voting our cohesive-obstruction variant
encodes, and \citet{capoccia2005} documents how interwar extremist parties
used parliamentary presence to deadlock the institutions they opposed. The
contrast between the two variants thus maps onto a substantive empirical
question about any given fragmented parliament: is its opposition composed
of office-seeking parties that bargain issue-by-issue, or of anti-system
blocs that obstruct? Our model predicts opposite passage regimes for the two.

This framing also makes a prediction that a pure
``fragmentation-collapse'' framing does not: under polarisation, the same
institutions that suffered the fragmentation collapse should benefit most
from discipline. Table~\ref{tab:ablation-polarized} confirms this.

\subsection{The passage--representation tradeoff as a single spectrum}

Figure~\ref{fig:tradeoff} and Table~\ref{tab:tradeoff-polarized} make a
concrete claim about the tradeoff between legislative throughput and
representational fidelity: the two quantities lie on a single axis for the
four institutions studied. Parliamentary sits at one end (high throughput,
low filter); republican at the other (low throughput, high filter);
semi-presidential variants between them. The mechanism is
straightforward. Discipline-plus-majority passage, the parliamentary
mechanism, has no ideological test: whatever the coalition proposes
passes. The presidential veto, the republican mechanism, \emph{is} an
ideological test: bills far from the executive's ideology are filtered
probabilistically. Committees add filtering in both but the effect is
smaller than the presidential veto.

This casts the parliamentary ``advantage'' in baseline passage rates
differently. Parliamentary systems pass more bills in absolute terms but
do not necessarily pass \emph{better} bills in the sense of matching
constituent preferences. The comparison requires joint consideration of
both axes, not rank-ordering on one.

\subsection{Policy and institutional-design implications}

The conventional response to fragmented-parliament gridlock is
\emph{electoral reform}: raise thresholds, move toward first-past-the-post,
encourage larger parties. Our results suggest a second route:
\emph{discipline reform}. Allowing free votes on more bills, weakening
leadership whip powers, or restructuring MP incentives toward their
districts rather than their parties could restore passage capacity under
fragmentation without touching electoral rules. Whether such reforms also
damage representation is an open empirical question: the parliamentary
filter is already weak in our model, so weakening discipline further might
spread passed bills without the presidential-veto compensation. This model
can illuminate that question but cannot yet answer it directly.

For new constitutional design, the Phase~F robustness result suggests a
tradeoff between fragility and filtering. Pure parliamentary maximises
passage at baseline but has a failure mode under fragmentation;
president-parliamentary partially insulates against that failure mode but
sacrifices some passage at baseline; republican is robust across all
scenarios but pays a throughput tax. Premier-presidential is closer to
parliamentary than its name suggests and shares the fragmentation
vulnerability.

\subsection{Relation to prior work}

\citet{linz1990} argued parliamentary systems outperform presidential
systems in producing governments and policy. Our baseline results
($0.742$ vs.\ $0.450$) echo that finding on passage rates, but our
fragmentation results reverse it: parliamentary collapses while presidential
holds steady. The conditional reversal recovers a pattern Linz's categorical
claim cannot predict. \citet{shugart1992} argued semi-presidential variants
should differ from each other in specific ways; we confirm that
president-parliamentary's partial rescue of passage under fragmentation is
a direct consequence of the president-driven formation rule. \citet{tsebelis2002}
argued additional veto players reduce policy mobility; our tradeoff finding
refines this by showing that veto players also improve representation
quality, so the choice is not ``veto players good or bad'' but ``veto
players trade throughput for filtering.''

\section{Limitations}
\label{sec:limitations}

The model is a stochastic pipeline model of institutional rules, not a
dynamic population model: legislators, parties, and constituencies are
static after initialisation, there are no elections, and no agent learns or
adapts. All institutional differences therefore reflect rule-level filtering
of an exogenous bill stream, and nothing here speaks to how representation
evolves under electoral feedback; adding endogenous selection (electoral
deselection of mavericks, party splits, constituency updating) is the
natural next step and would make discipline strength an outcome rather than
a parameter.

The hung-parliament behaviour flag resolves a genuine ambiguity rather than
a measured parameter: real fragmented chambers mix obstruction and
issue-by-issue bargaining, and the two pure regimes we compare are bounds,
not descriptions of any single chamber. The shipped default (cohesive
obstruction) reproduces the collapse regime reported in the comparative
literature, but the personal-vote null is the more conservative reading and,
in our view, the better baseline for future work.

The model treats legislator ideology as static and does not include
learning or adaptation. It has no endogenous party emergence, realignment,
or exit; the party structure is fixed at initialisation. It has no
bicameralism, judicial review, federalism, or presidential agenda-setting
powers beyond the veto. Bills have no path dependence: each bill is an
independent draw. We do not model pre-parliamentary agenda control (e.g.,
cabinet screening of government bills before introduction), which is why
our parliamentary baseline rate of $74\%$ is below the empirical UK and
German benchmarks of $90$--$95\%$ for government bills.

Cohabitation is detected in our semi-presidential runs (shares of
$12$--$26\%$ across scenarios) but does not produce meaningful passage
effects: our dismissal rate of $5\%$ per bill under president-parliamentary
is too gentle to create sustained gridlock, and the premier-presidential
veto is too soft to differentially block bills under
cohabitation. Future work should treat cohabitation dynamics more carefully,
including cabinet-to-legislator ideological alignment and veto thresholds
that shift under divided government.

The sensitivity analysis was run primarily on the baseline scenario; a
secondary run on the fragmented scenario (\S\ref{sec:results:sensitivity}) confirms
the \texttt{num\_parties} dominance but is not a substitute for per-scenario
sweeps of all parameters. The Morris and Sobol samplers include
\texttt{num\_parties} and \texttt{num\_constituencies} in the sweep space;
scenario definitions are therefore overlapping with the sensitivity space,
which is a methodological choice to allow continuous interpolation rather
than a discrete scenario lookup.

\section{Reproducibility}
\label{sec:reproducibility}

All results are deterministic under seed. The repository
\url{https://github.com/tofuadmiral/institutional-representation-abm} contains
the full code, $80$ automated tests (Python 3.13 on Ubuntu, GitHub Actions
CI), a pinned regression fixture, a Streamlit interactive companion
application, and a CoMSES metadata card. The command
\begin{verbatim}
python -m experiments.multiseed_comparison --scenarios all --seeds 200
\end{verbatim}
reproduces Figure~\ref{fig:forest}.
Table~\ref{tab:hung} and Figure~\ref{fig:hung} are produced by
\texttt{experiments.hung\_parliament};
Tables~\ref{tab:ablation-fragmented} and \ref{tab:ablation-polarized}
require additional invocations of \texttt{experiments.ablation};
Figure~\ref{fig:robustness} requires
\texttt{experiments.discipline\_robustness}; and
Figure~\ref{fig:tradeoff} requires \texttt{experiments.representation}.
All entry points are documented in the repository \texttt{README.md}.
Full runtime is under five minutes on eight cores. An interactive
Streamlit companion (\texttt{streamlit run streamlit\_app/app.py})
exposes every tunable configuration parameter as a slider and reproduces
the scenario-comparison, parameter-sweep, and ablation workflows used in
\S\ref{sec:results}.

\appendix

\section{ODD Protocol}
\label{app:odd}

The full ODD protocol \citep{grimm2010, grimm2020} is available in the
repository at \texttt{docs/ODD\_PROTOCOL.md}. Core elements are summarised
in \S\ref{sec:model}.

\bibliographystyle{plainnat}
\bibliography{references}

\end{document}